\documentclass[aps,prc,twocolumn,showpacs]{revtex4}

\usepackage{longtable}
\usepackage{graphicx}
\usepackage{epsfig}
\usepackage{dcolumn}
\usepackage{bm}

\begin{document}


\title{New nuclear structure features in trans-actinide nuclei.
}

\author{D. Bucurescu $^*$ and  N.V. Zamfir $^\dagger$} 
  
\affiliation
{Horia Hulubei National Institute for Physics and Nuclear Engineering, 
RO-077125 Bucharest-M\u agurele, Romania}

\email[]{bucurescu@tandem.nipne.ro}
\email[$^\dagger$]{zamfir@tandem.nipne.ro}


\date{\today}

\begin{abstract}
The structural evolution of the heavy nuclei, with $Z > 82$, is investigated by 
looking at 
the differential variation of the two-neutron separation energies. 
It indicates, by non-monotonous behavior at certain 
neutron numbers, structure 
phenomena such as major shell ($N = 126$) and deformed subshell ($N = 152$) closures. 
Another interesting effect is observed at $N \sim 142$, which is very well 
correlated with a previously observed, intriguing behavior of quantities measured in 
alpha decay, such as relative branching ratios and hindrance factors of excited states
from the ground state band of deformed nuclei in this region. Corroboration of the
existing experimental data indicates another possible deformed subshell closure.  
\end{abstract}


\maketitle

\section{Introduction}

The heavy nuclei with $Z > 82$ are generally less accessible for detailed 
studies with usual nuclear spectroscopy methods. For example, gamma-ray spectroscopy studies 
of heavier nuclei in this region began to occur only recently  (see e.g. \cite{250Fm}).
On the other hand, alpha decay represents one of the main tools for the spectroscopy of 
excited states of nuclei in this region. 
However, we have recently shown that 
the population of the excited states through alpha decay (the so-called fine structure) 
is not thoroughly understood, precluding the full use of this process as a spectroscopic
tool \cite{alpha0,alpha}. Thus, an intriguing evolution was observed  
for both the relative branching ratios and hindrance factors of the $2^+$, $4^+$ and $6^+$ 
states from the ground state band of deformed nuclei in the Uranium to Californium isotopes,
which is apparently related to intricate nuclear structure details. 
Figure 1 illustrates one of the most conspicuous observed effects, a remarkable maximum around
mass 240 ($N$ values around 142) in nuclei above Uranium, in the relative branching ratio
and hindrance factor of the $4_1^+$ state. Such a behavior has been rather unexpected in
these nuclei which are considered well deformed rotors, and its explanation is still 
pending, as it could not be correlated 
with any of the evolutions of several investigated nuclear structure 
observables \cite{alpha0,alpha}.

Various features of the nuclear structure can be deduced by observing the 
evolution of different quantities that characterize the nucleus. Thus, for 
example, most nuclear structure observables, even the simplest ones, such as, in 
the even-even nuclei,
the energy of the first excited $2^+$ state, or the electromagnetic transition 
probability of this state to the ground state, 
show clear indications for the magic numbers (major shell closures), and also 
display characteristic 
evolutions in between these numbers (the collective features).
It is also known that the nuclear masses concentrate valuable information on the 
structural changes. The variation of the nuclear mass with
the number of protons/neutrons shows many irregularities that can be related 
to different structural changes, the strongest being those due to the 
magic numbers.   
Finer details of the structure evolution can be 
observed by looking at the evolution of the nuclear masses through a  
proper 'eye-glass'. Such a  systematic investigation was performed in Ref. \cite{S2n}
by looking at the evolution of a mass-related quantity, the  two-neutron separation energy
$S_{2n}$ in the even-even medium mass nuclei (with $Z$ from 28 to 80), and using as
eye-glass the differential variation of this quantity  as a function of
$N$ and $Z$. Various other differential observables were 
used in Ref. \cite{cakirli} to highlight evolutions and changes in  
nuclear structure.  

Because the approach based on the differential of the two-neutron separation energy
proved very useful in indicating not only major shell closures 
but also different subshell closures as well as points of phase/shape transitions, 
we decided to investigate  in the same way the evolution of the known alpha-decaying 
nuclei above $Z = 82$. As a result, 
we have found interesting features in the evolution of the derivative of $S_{2n}$ with 
respect to $N$ and also unexpected empirical correlations between the variation of this 
derivative and those of the alpha-decay fine structure quantities. 
The analysis of the existing set of experimental data points to the importance 
of deformed subshell closures in these nuclei.

\section{Systematics of the two-neutron separation energies}

The two-neutron separation energies $S_{2n}$ were taken from the recent mass
evaluation, Ref. \cite{ame2012}.  
Figure 2(a) shows the variation of $S_{2n}$ as a function of the number of neutrons
$N$ for the isotopes with $Z$ from 82 to 104 (Pb to Rf). 
To show the trends also for the isotopic 
chains with the highest $Z$ (102 and 104)  we included in some cases 
estimated values and errors given in  \cite{ame2012} (these cases can 
be recognized by their larger error bars). 
A feature that is common to all mass regions 
\cite{S2n} is that $S_{2n}$ decreases monotonously with increasing $N$, as in Fig. 2(a), 
the curves for 
different isotopes being almost parallel; a more accentuated decrease takes place after the
magic numbers (it is more difficult to remove two neutrons from a nucleus with a magic neutron 
number). This accentuated variation is easily observed in graph (a) at $N = 126$, the well known 
spherical major shell closure. 
Finer details can be obtained by looking at the differential variation of 
$S_{2n}$ with $N$. This can be defined as:
\begin{equation} dS_{2n}(Z,N) = [S_{2n}(Z,N+2) - S_{2n}(Z,N)]/2 \end{equation}

From the general properties of $S_{2n}$ it follows that for a certain mass region, 
such as that
shown in Fig. 2, $dS_{2n}$ is negative and approximately constant (actually, a weak, 
practically linear variation with $N$ is observed \cite{S2n}), except for some \
regions with a more accentuated variation.
A sudden jump to much larger 
negative values is observed for $dS_{2n}(Z,N)$ at the magic number $N = 126$  - graph (b). 
 The evolution of $dS_{2n}(Z,N)$ shows other interesting structure effects
as well. For example, in the Ba to Gd nuclear region there is another obvious deviation from the 
smooth behavior, a positive 
bump at $N = 88$ -- 90 \cite{S2n}, which corresponds to the critical 
point $X(5)$ of
the shape/phase transition between spherical and deformed nuclei. Similar features were observed in other nuclear regions, 
e.g., in the Sr, Zr, Mo nuclei a negative kink in $S_{2n}$ (more reduced than the one at
$N = 50$) which corresponds to the known $N = 56$ subshell closure, and a positive one at $N = 60$, 
where a sudden onset of deformation is known, that can be also associated 
with an $X(5)$ critical
point transition. 
Indications were also found for different other weak subshell closures
\cite{S2n}. In the nuclear region considered here there is
another clear negative bump, of smaller amplitude,  at $N=152$. Nuclei in the region $Z =100,~
N = 152$ are all deformed, showing clear rotor features, therefore it is
a deformed subshell closure. The 'magic' character of $N = 152$ has been known for a long time.
Various calculated Nilsson diagrams show a deformed gap at 
$N = 152$ (e.g., Ref. \cite{gustafson}). 
This shell closure was shown to have a marked effect on the spontaneous fission half-lives of 
those nuclei \cite{Leino}, and its effects on the structure of the nuclei 
are also seen both in experimental data and calculations, see e.g. 
\cite{chatillon,sobiczewski,patyk}. 
One should emphasize the fact that a 
deformed subshell
gap shows up in $dS_{2n}$ in the same way as the  spherical 
major shell closure, that is, 
as a negative  bump. 
In addition to the two 'shell-closure' negative bumps clearly observed 
in Fig. 2(b), there is a smaller-scale negative bump at $N = 142-144$ for the isotopes of U to
Cf. Fig. 3 shows a larger
scale detail of graph 2(b), where it can be seen that for the U, Pu, Cm, and maybe Cf isotopes, 
there is indeed a negative bump 
that can be clearly distinguished from the fluctuating  'background' 
of $dS_{2n}$ in this region. 

Similar conclusions are drawn from examining another quantity related to the nuclear masses, 
the alpha-decay $Q$-value, $Q_{\alpha}$. We take this quantity from the same tables \cite{ame2012}
and Fig. 4 presents the variation with $N$ of both $Q_{\alpha}$ and its differential 
$dQ_{\alpha}(Z,N) = [Q_{\alpha}(Z,N+2) - Q_{\alpha}(Z,N)]/2$, for the same nuclei from 
Fig. 1. Again, the differential $dQ_{\alpha}$ clearly shows specific bumps (positive, this
time) at the 
$N = 126$ spherical shell closure and at the $N = 152$ deformed subshell closure, as well as  
a smaller  bump at $N = 142-144$, which is statistically significant for the higher $Z$ values
(above 92).

While this effect at $N \sim 142$, observed in both $dS_{2n}$ and $dQ_{\alpha}$ is smaller 
than that due to the 
$N = 152$ subshell closure ($\sim$150 keV deep, compared to $\sim$400 keV), we cannot 
overlook it because, as shown below, it is a real effect 
that clearly shows up in the evolution of other nuclear structure observables.

\section{Correlation between $dS_{2n}$ and alpha-decay fine structure observables}

As already emphasized, details concerning the absolute 
rates of the  population of excited states in the ground state band  of even-even 
trans-lead nuclei also constitute valuable indication on their structure.  
In Refs. \cite{alpha,alpha0} we  have examined the
evolution of the following quantities: the branching ratios $B_{r,i}$ for the population of
the excited states '$i$', and the hindrance factors $HF_i$. The hindrance factors are 
defined as \cite{alpha}: 
\[ HF_i = \frac{B_{r,gs}}{B_{r,i}} \frac{P_i}{P_{gs}} \]
where  $gs$ denotes the $0^+$ ground state, and $P$ are the corresponding penetrabilities 
of the alpha particle through the barrier.  
 In particular, for the $4^+$ state of the gsb, a
pronounced  maximum was observed for both these quantities, corresponding to the U, Pu, and 
Cm nuclei with $N$ values of 142 -- 144 (see Fig. 1). 
This striking maximum, observed in the middle of
a region of well deformed nuclei could not be 
correlated with the behaviour of any of the different simple structure indicators that were 
investigated: the energy of the $2_1^+$ state, the ratio $R(4/2) = E(4_1^+)/E(2_1^+)$, 
the quadrupole ($\beta_2$) or hexadecapole ($\beta_4$) deformation parameters, 
or the $J_0$ and $J_1$ parameters of the fit of the gsb energies with the VMI formula
(or Harris parametrization) \cite{VMI}, $J_0$ being close to the moment of inertia and  $J_1$
characterizing the stiffness, or rigidity, of the nucleus \cite{alpha}. 

It turns out that the evolutions of both  $B_{r,gs}/B_{r,i}$ and $HF_i$ for the 
$4^+$ state of the gsb (Fig. 1) are well correlated with that of the $dS_{2n}$
quantity discussed above.  In the data discussed below, the hindrance factor values 
are those adopted in the ENSDF database \cite{ENSDF}, which correspond to penetrabilities 
calculated according to the recipe of Preston \cite{Preston}.
As discussed in \cite{alpha}, hindrance factors calculated with different methods 
differ only in absolute values but their evolution is very similar. 
Figure 5 shows the evolution of all these quantities as a function of the neutron number $N$.
The symbols represent the isotopes of 
Ra, Th, U, Pu, Cm, and Cf for which the branching ratio of the $4^+$ state was
measured. 
We have noticed that there is a good similarity between the variation with $N$ of 
-$dS_{2n}(Z,N)$ and that of the alpha decay quantities, provided 
the $N$-values are shifted by two units for the latter: that is, in Fig. 5(b) and (c), at 
the x-axis corresponding to the neutron number $N$  are 
represented the relative  branching ratio and the hindrance factor for the daughter nucleus
$(Z,N+2)$. 
%
%
Figure 5 shows a rather good correlation between -$dS_{2n}(Z,N)$
and the logarithms of the $B_{r,gs}/B_{r}(Z,N+2)$ and $HF(Z,N+2)$ values of the $4^+_1$ state.  
The correlation is  better in the case of the 
hindrance factors (quantities in which the barrier penetrability was corrected for). 
By comparing graphs (a) and (c) in Fig. 5 one can see that there 
is a very good similarity for the variation with $N$ of the two quantities,
 with the exception of the Plutonium nuclei, which have 
hindrance factors that are larger than those suggested by the superposition
of the two graphs. 
  
  The quality of the correlation between $-dS_{2n}(Z,N)$ and the hindrance factor
of the $4^+$ state of the $(Z,N+2)$ nuclei 
can be better appreciated in Fig. 6. 
All nuclei in this graph show a reasonably good linear correlation. 
 There is some scattering of the data around this straight line, one of
the most deviating points being that of $^{250}$Cf (daughter nucleus), nevertheless the graph 
demonstrates a reasonably good correlation of the two quantities. 
As noted above, the Plutonium isotopes jump out of this general pattern, at higher $HF$
values, but their evolution is still well represented by a straight line with the same
slope as that for the other nuclei.
One should note that these nuclei, where the alpha decay branching is known 
for the $4^+$ state  correspond mainly to those from the region where 
$dS_{2n}$ presents the small negative bump at $N \sim 142$
discussed  in the previous section. Therefore, the structure effect seen in $dS_{2n}$ 
manifests itself strongly in a nuclear structure observable, namely the $4^+_1$ state 
alpha decay hindrance factor. 

As discussed in Ref. \cite{alpha},  the relative branching ratio and the hindrance factor values 
corresponding to the $6^+_1$ excited state vary, as functions of $N$, practically out of
phase with those of the $4^+_1$ state (Fig. 2 in \cite{alpha}), therefore they are also 
reasonably well correlated 
with $dS_{2n}$. Because in this case there are less data points, and their scattering is 
somewhat larger \cite{alpha} , we concentrate the discussion on the $4^+$ state data. 
%

\section{Discussion}

In the following we discuss   the possible meaning of the small bump seen in the 
variation of $dS_{2n}$ (Figs. 2 and 3) at $N \sim 142$. 
The isotopes with $Z$ = 92 to 98 show a clear minimum at $N = 142-144$, which is about 
150 keV deep below  the average background.   
As it is a negative bump, just like the other two 
corresponding to (sub)shell closures at $N = 126$ and 152, respectively, it is tempting to 
attribute it to another deformed subshell closure as well. Actually, 
for the Cm isotopes the minimum at $N = 144$ has practically the same depth with that 
at $N = 152$, and the Cf isotopes show a rather similar evolution.
A good argument in favor of assigning this minimum to a subshell closure 
 would be to compare the behavior of HF($4^+$) in this region of $N$ values
with that of the same quantity at a well established shell closure. 
Unfortunately, the alpha decay
experimental data are very scarce in this respect. Figure 7 presents 
a graph similar to that in Fig. 5, for all the other cases
where the $4^+$ state was measured in alpha decay, in nuclei close to $N=126$. Only
one Pb isotope and three Rn isotopes (as daughter nuclei) were found in this region. 
Figure 7(a) shows the strong, sharp maximum of the $-dS_{2n}(Z,N)$ quantity
at the magic number  $N=126$, and at larger $N$ an almost constant value
smaller by a factor of about 15. 
Both the relative branching and the $HF$ for the $4^+_1$ state 
of the $(Z,N+2)$ nuclei behave, as a function of $N$,  similarly  to
 $-dS_{2n}(Z,N)$,  also having a large 
value at $N=126$ (that point correponds to the daughter nucleus $^{210}$Pb), 
and a smaller average value at higher $N$.
 For nuclei with $Z$ around 100, which feel quite well 
the $N=152$ deformed shell closure (see Fig. 3) there are no
measured data for the branch of the $4^+$ state. 
Based on these (unfortunately limited in number) experimental data, one may conclude,
with some caution,  
that the signature of a neutron shell closure is a  maximum in the value of the
relative branching ratio, or of the hindrance factor, of the $4_1^+$ state. 
By analogy, one can then suggest that the intriguing maximum in HF($4^+$) (e.g., Fig. 5)
indicates a possible subshell closure showing up in Figs. 2 and 3 
as a bump at $N = 142-144$,
for the isotopes with $Z = 92$ to 98.   

Until now, experimental effects indicating such a neutron subshell closure have not been 
discussed in the literature. All nuclei corresponding to this bump have large 
quadrupole deformations, $\beta_2$ larger than 0.25 \cite{Raman,alpha0}, therefore this
should also be a deformed subshell closure. On the other hand, different calculated 
Nilsson single-particle level schemes indicate that such a gap may 
occur for $N = 142$ at $\beta_2$ values around 0.22 -- 0.23, in addition to the 
$N = 152$ gap occuring at higher values, with $\beta_2$  around 0.28 -- 0.29 
\cite{gustafson,meldner}. In some cases, the subshell gap predicted 
at $N = 142$ in nuclei of mass $\approx 240$ is explicitely
indicated, in addition to that at $N = 152$ 
\cite{Nilsson,Ragnarsson,book}. 
The occurence
of such a shell gap should result from intricate structure details that dictate the relative
positions of the deformed orbitals originating in the spherical shells 
$1i_{11/2},~1j_{15/2}$, and 2$g_{9/2}$.  
We finally note that the existence of another, deformed subshell closure at $N \sim 142$ 
in addition to that at $N = 152$
is also consistent with the observed variation of the rigidity of these deformed 
nuclei. In ref. \cite{alpha0,alpha} we noted that the nuclear stiffness coefficient (which
is proportional to the inverse of the $J_1$ Harris parameter \cite{VMI}) 
steadily increases  with mass for the deformed nuclei that are 
discussed here. The explanation could be that after achieving 
permanent deformation 
(generally in nuclei with $N$ just below 140), one passes through the two regions of subshell
closure, where the stiffness increases due to the  reducing of the effect of the 
pairing correlations.  

\section{Conclusions}
In summary, we have investigated the systematics of both the differential variation of the 
two-neutron separation energies $S_{2n}$ and of the alpha decay 
branching ratios and hindrance factors of excited states in deformed trans-lead nuclei. 
The differential $dS_{2n}$ proves a very sensitive tool for 
disclosing important information 
on nuclear structure changes in these deformed nuclei. Thus, besides the well-established 
$N = 126$ magic
number, it clearly shows the magic character of $N=152$, where there is a deformed subshell
closure. A weaker effect is visible in  $dS_{2n}$ as another negative bump 
at $N \sim 142$, which occurs for the isotopes with $Z = 92$ to 98.
This bump is rather well correlated with the extrema occuring in 
the behaviour of the alpha decay hindrance factors for the 
$4_1^+$ and $6_1^+$ states in the daughter nuclei (a maximum and a minimum, 
respectively)  \cite{alpha}. 
By analogy of the behavior of this correlation with that for nuclei around the $N=126$ shell gap, 
one can tentatively propose that the small bump indicates another (weaker)  
deformed subshell closure, which was actually predicted by Nilsson model
calculations.
It is interesting to observe (Fig. 3) that the Cm isotopes, for which the 
experimental data span 
the $N$ range from 140 to 152, 
'feel' equally strong both the $N = 144$ assumed subshell gap and 
the well-recognized $N = 152$ gap.  
An experimental confirmation of the observed type of 
correlation (larger $HF$ values at the 'magic' numbers) also for the heaviest nuclei
 with $Z \ge 96$ spanning the $N=152$ subshell closure would
be extremely valuable. 
We have no explanation why the $N \sim 142$ bump 
in $dS_{2n}$, clearly indicating an interesting nuclear structure effect, 
manifests itself strongly 
in the alpha decay population of the higher spin members of the ground state bands 
$4_1^+$ and $6^+_1$. It does not appear to have a visible
effect on the 
$2_1^+$ state data \cite{alpha}. One should remark, nevertheless,  that the $2_1^+$ state 
is populated in the alpha decay with a relative branching ratio $B_{r,gs}/B_{r,2^+}$  roughly
in the range 2 to 10, whereas the population of the $4_1^+$ state appears, by
comparison, as a much weaker phenomenon, 
with a relative branching ratio 
roughly between 100 and 4000 (Fig. 4), and even larger, 
of the order of 
$10^6$ at the magic number 126 (Fig. 7). 


The possible occurence of another deformed subshell gap at $N \sim 142$ 
opens up interesting questions about the intricate structure details of these nuclei, 
because the equilibrium shapes of such massive nuclei 
result from the interplay of many degrees of freedom, 
among which the quadrupole, hexadecapole, and octupole deformations play an important 
role. Also, how subshell closures in $N$ correlate with $Z$-values is another 
interesting aspect. 
Finally, although the detailed mechanism of the population of the excited states 
in the alpha decay process is not understood yet, the present experimental systematics 
indicate that the alpha decay fine structure is a spectroscopic tool with an extraordinary 
sensitivity, comparable to that of  the 'eye glass' constituted by the differential 
variation of the two-neutron separation energy.



\begin{acknowledgments}
 We acknowledge partial
support from UEFISCDI-Romania
under program PN-II-IDEI, Contract No. 127/2011.
\end{acknowledgments}

\newpage

\begin{figure}[htbp]
\epsfig{figure=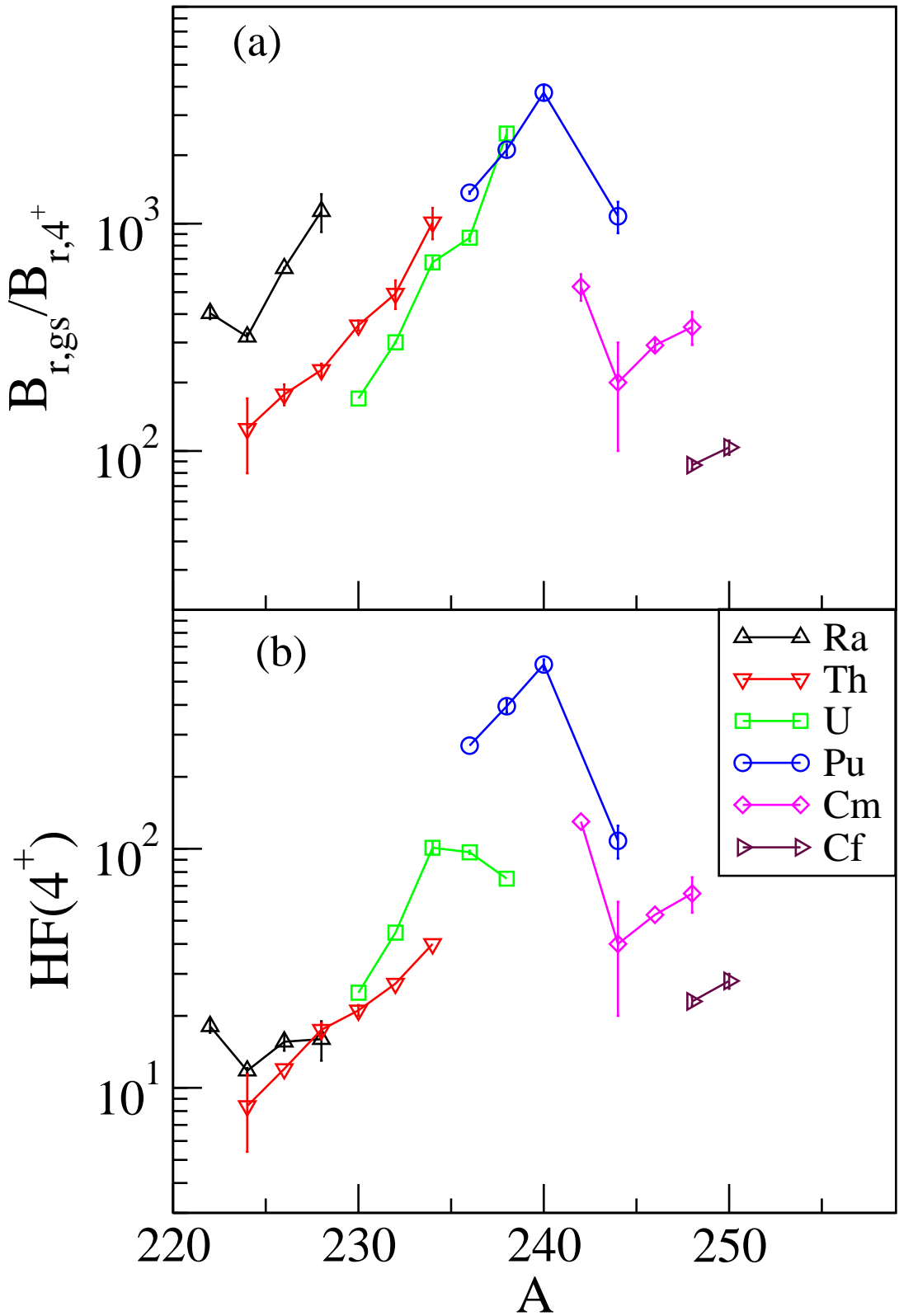,width=0.8\textwidth,angle=0}
\caption{(Color online) Variation of the the experimental alpha decay relative branching ratios and 
hindrance factors with the mass number, 
for the $4_1^+$ excited state in isotopes with $Z = 88$ to 98 (as presented and 
discussed in Refs. \cite{alpha0,alpha}).} 
\label{fg1}
\end{figure}

\begin{figure}[htbp]
\epsfig{figure=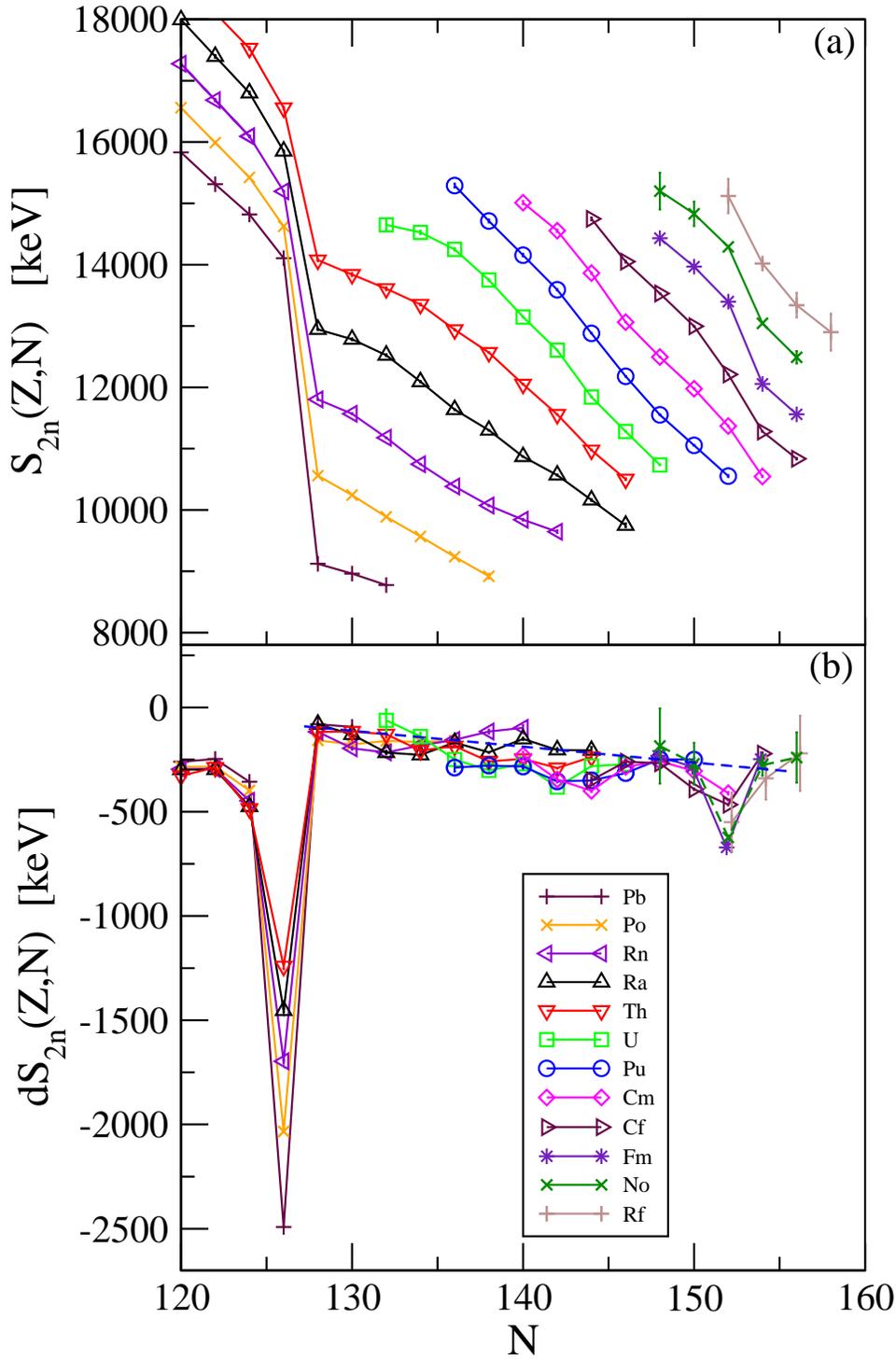,width=0.7\textwidth,angle=0}
\caption{(Color online) Evolution of $S_{2n}$ (a) and $dS_{2n}$ of eq. (1) (b)
for the isotopic chains Pb to Rf.  The average monotonous variation of $dS_{2n}$ is 
indicated by a dashed straight line drawn by hand.} 
\label{fg2}
\end{figure}

\begin{figure}[htbp]
\epsfig{figure=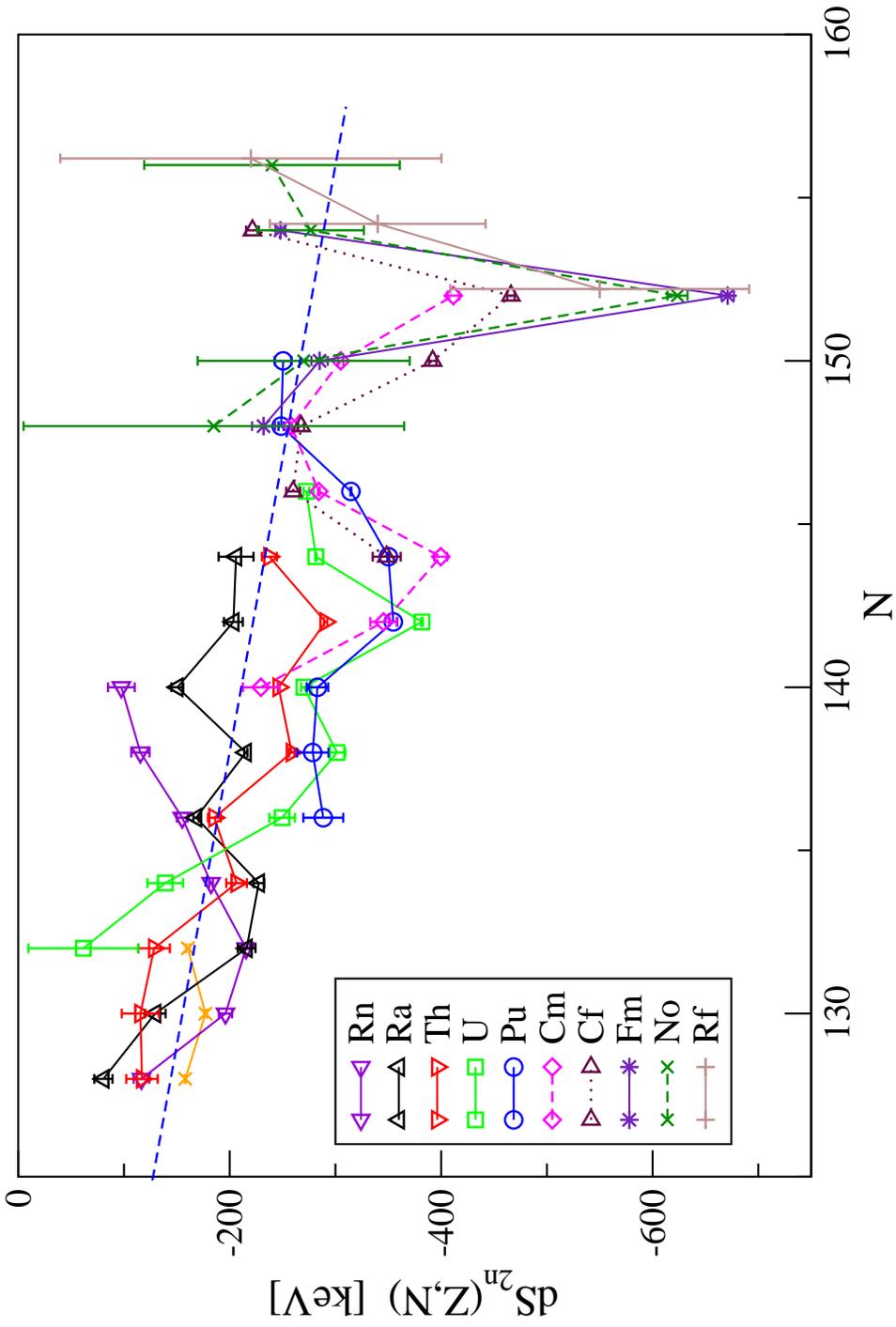,width=0.7\textwidth,angle=0}
\caption{(Color online) Detail of graph (b) from Fig. 2.} 
\label{fg3}
\end{figure}

\begin{figure}[htbp]
\epsfig{figure=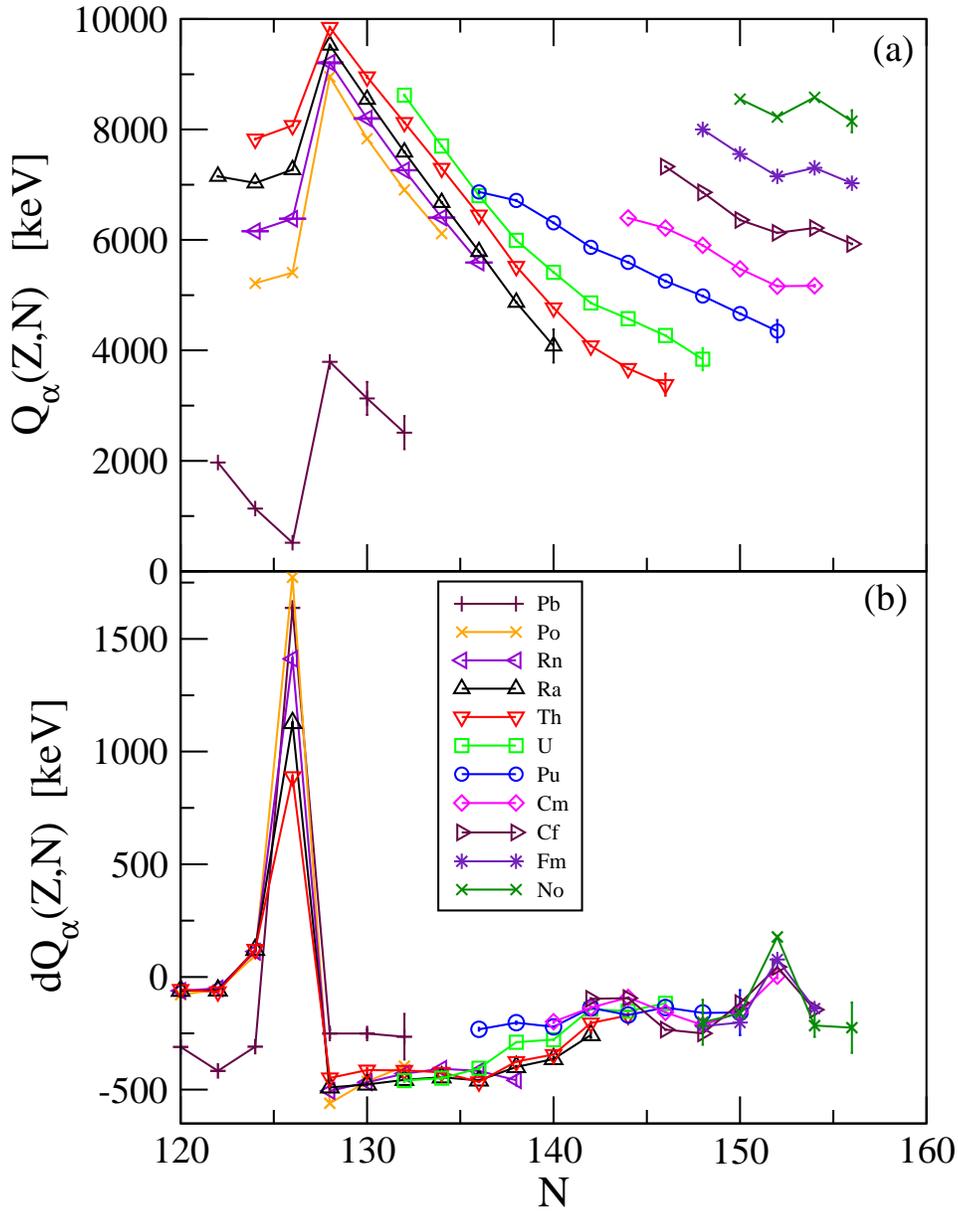,width=0.7\textwidth,angle=0}
\caption{(Color online) Behavior of the $\alpha$-decay $Q$-value $Q_{\alpha}$ (a), and its  
differential variation $dQ_{\alpha}$ defined in the same way with $dS_{2n}$ of eq. (1), 
$dQ_{\alpha}(Z,N) =
[Q_{\alpha}(Z,N+2) - Q_{\alpha}(Z,N)]/2$ (b), 
for the same Pb to No isotopic chains of Fig. 2. } 
\label{fg4}
\end{figure}

\begin{figure}[htbp]
\vspace*{0mm}
\epsfig{figure=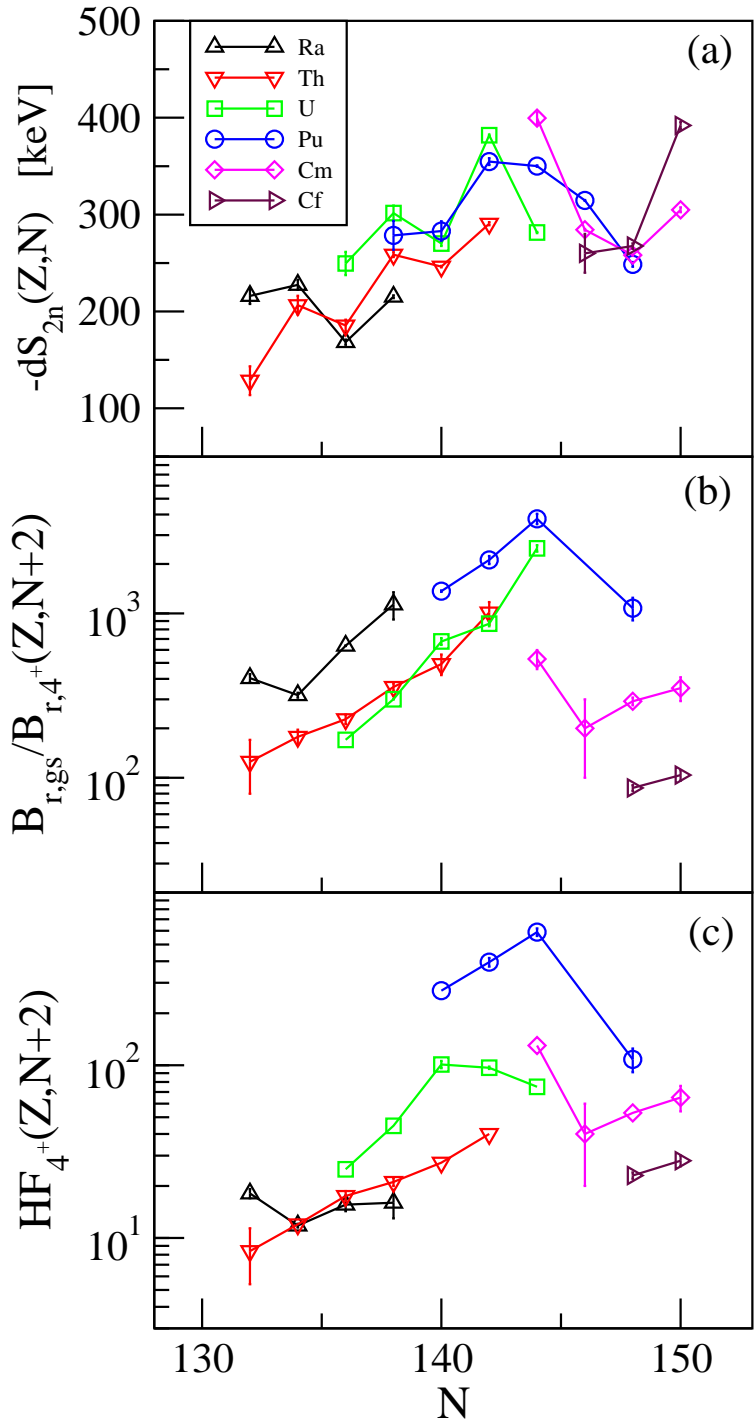,width=0.55\textwidth,angle=0}
\caption{(Color online) Comparison of the variations with the number of neutrons 
of the differential 
quantity $dS_{2n}$ (a), and of the $\alpha$-decay
relative branching ratios 
$B_{r,gs}/B_{r,4^+}$ (b) and 
hindrance factors  $HF(4^+)$ (c) for the $4^+$ state of the ground 
state band in the indicated daughter isotopic chains. As discussed in text, by representing 
the later two quantities for the $(Z,N+2)$ daughter nuclei as a function of $N$, one obtains 
the best similarity of graphs (b) and (c) with graph (a).} 
\label{fg5}
\end{figure}

\begin{figure}[htbp]
\vspace*{9mm}
\epsfig{figure=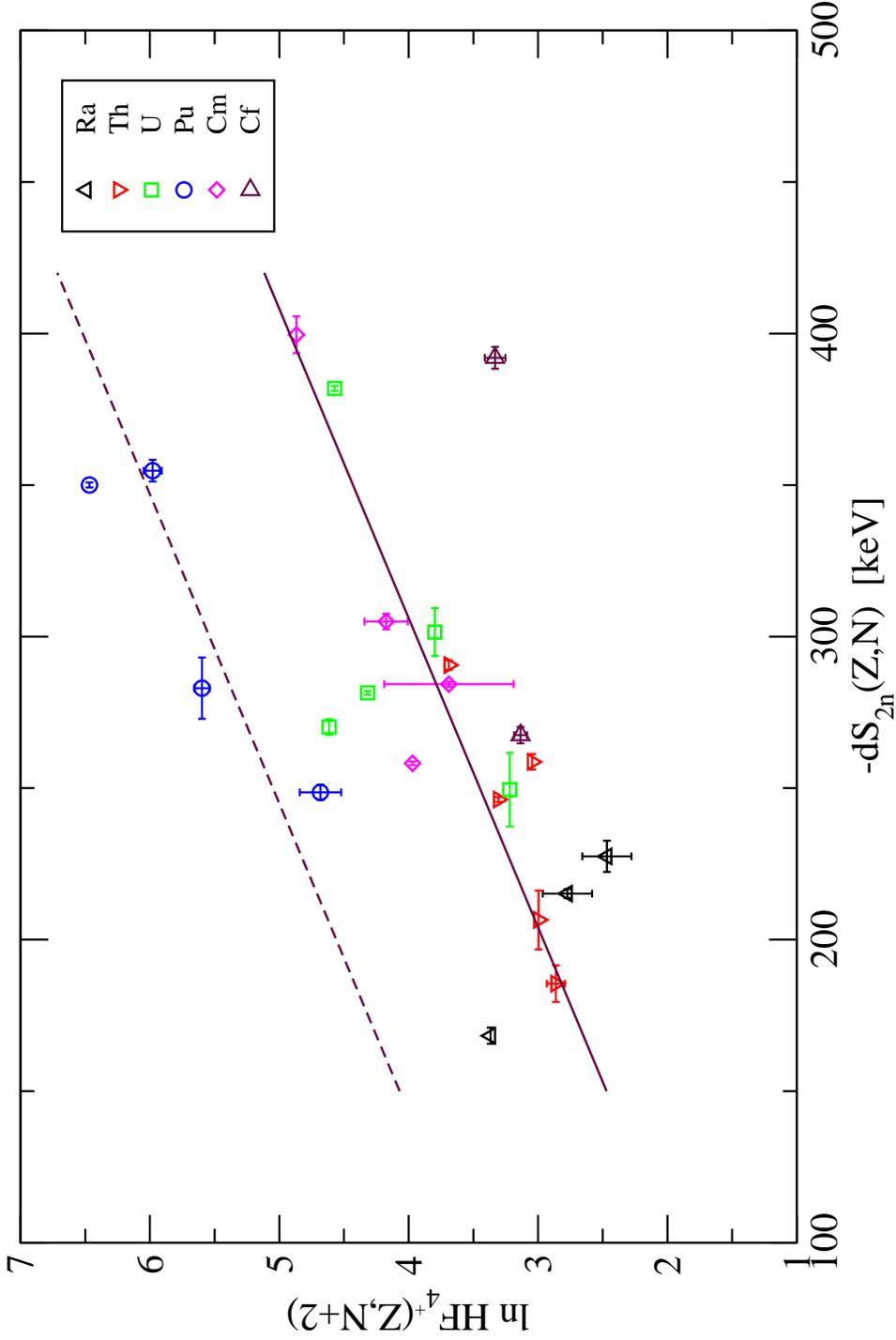,width=0.7\textwidth,angle=0}
\caption{(Color online) Correlation between the logarithm of the hindrance factors
(Preston definition) of the $4_1^+$ state in the $(Z,N+2)$ daughter nuclei
and $-dS_{2n}(Z,N)$ defined by eq. (1). The continuous curve is a linear fit to the data 
(to all shown nuclei, except the Pu isotopes), with a correlation coefficient of 0.85; 
the dashed line has an identical slope and 
was drawn such as to pass through the Pu isotope points. } 
\label{fg6}
\end{figure}

\begin{figure}[htbp]
\vspace*{9mm}
\epsfig{figure=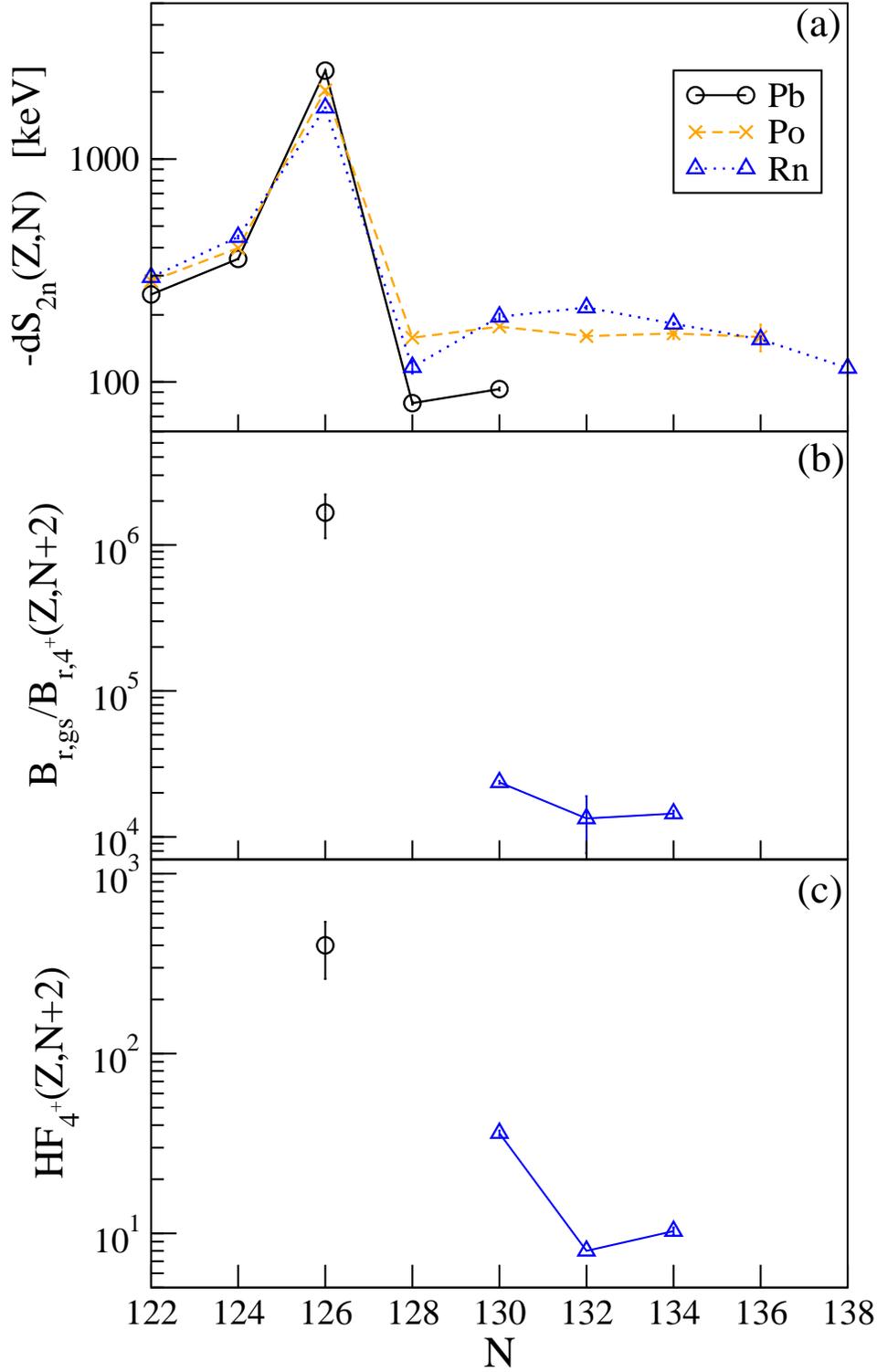,width=0.7\textwidth,angle=0}
\caption{(Color online) Similar to Fig. 5 but for Pb to Rn daughter nuclei.} 
\label{fg7}
\end{figure}





\end{document}